\documentclass[preprint,amsmath,amssymb,aps,prl,twocolumn,epsfig,showpacs,bibliography,lengthcheck]{revtex4-1}
\usepackage{CJK} 
\usepackage{color}
\usepackage{graphicx}
\usepackage{dcolumn}
\usepackage{bm}
\usepackage{hyperref}
\graphicspath{{../figures/}}

\begin{document}
\begin{CJK}{UTF8}{gbsn}
\title{Hyperfine anomalies in Fr: boundaries of the spherical single particle model}
\author{J. Zhang (张颉颃)$^{1}$}
\author{M. Tandecki$^{2}$} 
\author{R. Collister$^{3}$} 
\author{S. Aubin$^{4}$}
\author{J. A. Behr$^{2}$} 
\author{E. Gomez$^{5}$}
\author{G. Gwinner$^{3}$}
\author{L. A. Orozco$^{1}$}
\author{M. R. Pearson$^{2}$} 
\author{G. D. Sprouse$^{6}$} 
\affiliation{$^{1}$Joint Quantum Institute, Department of Physics, University of Maryland, and National Institute of Standards and Technology, College Park, MD 20742, U.S.A. 
$^{2}$ TRIUMF, Vancouver, BC V6T 2A3, Canada. 
$^{3}$ Dept. of Physics and Astronomy, University of Manitoba, Winnipeg, MB R3T 2N2, Canada
$^{4}$Department of Physics, College of William and Mary, Williamsburg VA 2319, U.S.A.  
$^{5}$Instituto de F{\'i}sica, Universidad Aut{\'o}noma de San Luis Potos{\'i}, San Luis Potos{\'i} 78290, M{\'e}xico. 
$^{6}$Department of Physics and Astronomy, Stony Brook University, Stony Brook, New York 11794-3800, U.S.A.}
\date{\today}

\begin{abstract}

We have measured the hyperfine splitting of the $7P_{1/2}$ state at the 100 ppm level in Fr isotopes ($^{206g,206m, 207, 209, 213, 221}$Fr) near the closed neutron shell ($N$ = 126 in $^{213}$Fr).  The measurements in five isotopes and a nuclear isomeric state of francium, combined with previous determinations of the $7S_{1/2}$ splittings, reveal the spatial distribution of the nuclear magnetization, i.e. the Bohr-Weisskopf effect. 
We compare our results with a simple shell model consisting of unpaired single valence nucleons orbiting a spherical nucleus, and find good agreement over a range of neutron-deficient isotopes ($^{207-213}$Fr). Also, we find near-constant proton anomalies for several even-$ N$ isotopes. 
This identifies a set of Fr isotopes whose nuclear structure can be understood well enough for the extraction of weak interaction parameters from parity non-conservation studies.
\end{abstract}

\pacs{21.10.Gv,27.80.+w,32.10.Fn}
\maketitle
\end{CJK}

Weak interaction studies in heavy atoms require for their interpretation precise knowledge of the atomic and nuclear wavefunctions. 
To extract nucleon-nucleon weak interaction couplings from the weak interaction induced parity-violating anapole moment \cite{wood97}, nuclei with simple and regular magnetic properties are desirable \cite{flambaum97, haxton02, wilburn98}. 
The nuclear magnetic moment is used to benchmark nuclear structure theories for calculating the anapole moment~\cite{haxton02}, which is a contact field effect produced inside the finite extent of the nucleus.
Here we explore the regularity of the magnetic properties of a chain of Fr isotopes and find that $^{207-213}$Fr in the vicinity of the neutron shell closure mark a range where the nuclear structure is sufficiently tractable for standard model tests  and constraints on new physics \cite{roberts14c}.

To lowest order, the atomic hyperfine interaction can be described using a point-like nucleus characterized by the magnetic dipole moment.
Deviations from the point-like approximation of the nucleus, referred to as hyperfine anomalies, come from considering how finite magnetic and charge distributions affect the interaction between the magnetization of the nucleus and the magnetic field created by the electrons. 
The magnetic contribution is known as the Bohr-Weisskopf  (BW) effect~\cite{bohr50,buttgenbach84}. The difference in the nuclear charge distribution (Breit-Rosenthal (BR) effect \cite{rosenthal32,crawford49,rosenberg72}) produces very small variations between isotopes, leaving the BW effect dominant \cite{martensson95,Note1BR}. 
As a new generation of proposed parity violation experiments in atoms (including Fr) and molecules starts \cite{tandecki13a, gomez07, leefer14, nunez13, bougas12, cahn14, roberts14b}, it is important to understand the limiting factors due to the nuclear structure, e.g. the nuclear magnetization, for the interpretation of parity-violating anapole moments \cite{haxton02, flambaum97, wilburn98}. 

Here we present a systematic study of the variations of the hyperfine splittings (HFS) in a chain of Fr isotopes. 
Our measurements include $^{213}$Fr with the closed neutron shell (magic number $N$ = 126), $^{206g,m}$Fr, $^{207}$Fr, and $^{209}$Fr on the neutron-deficient side, and the neutron-rich $^{221}$Fr. 
In the region of $^{207-213}$Fr with up to 6 neutron holes, we find near-constant magnetic hyperfine anomalies for the odd-$Z$, even-$N$ isotopes \cite{grossman99}. 
The neutron rich odd-even isotope $^{221}$Fr shows a different behavior due to the deformation of the nucleus. The odd-$Z$, odd-$N$ isotopes  have anomaly contributions from both the proton and the valence neutron. 

BW effect measurements usually require precise knowledge of both, hyperfine structure constants and magnetic moments. 
Measurements of the nuclear magnetic dipole moment in Fr are limited to $^{211}$Fr \cite{ekstrom86} and $^{210}$Fr \cite{gomez08}.  Empirical values for other isotopes are usually obtained by scaling with the isotopic ratios of the hyperfine constants based on these two experiments, both of which have uncertainties larger than 1\% and cannot be used to extract the hyperfine anomaly. 
A different approach to study the BW effect which circumvents the limited precision in the magnetic moments comes from the suggestion by Persson~\cite{persson98} that was implemented in Fr \cite{grossman99}, Tl \cite{barzakh12} and Eu~\cite{zemlyanoi10}. This method relies on looking at the ratio between hyperfine splittings of two levels where the electron wavefunctions are overlapping differently with the nuclear wavefunction, and consequently is sensitive to the differences in their hyperfine anomalies. 
We can normalize the change in this ratio to a specific isotope ($^{213}$Fr) for purpose of directly revealing the contribution from the neutron wavefunction. 
See Ref. \cite{buttgenbach84} for a review and Ref. \cite{persson13} for a recent compilation of all available hyperfine anomaly data.

The magnetic hyperfine interaction $W$ can be written as \cite{stroke61,persson98}
\begin{equation}
W_{\rm extended}^{l} =W_{\rm point}^{l}(1+\epsilon_{l}),
\end{equation}
where $\epsilon_{l}$ is a small quantity that depends on the particular isotope, and on the atomic state ($l = S$ or $ P$). The $7P_{1/2}$ electron overlaps  with the nuclear wavefunctions more uniformly than the $7S_{1/2}$ electron. The ratio $R$ of the hyperfine splittings for an isotope with mass number $A$ is sensitive to the nuclear magnetization distribution \cite{grossman99}
\begin{eqnarray}
R_{\rm HFS}(A) & = & \frac{ W_{\rm extended}^{S}}{ W_{\rm extended}^{P}} \approx R_{0}(1+\epsilon_{S}(A) -\epsilon_{P}(A)), ~~
\label{rho}
\end{eqnarray}
with $R_{0}$ the ratio of hyperfine structure constants for a point nucleus. Since both states have $J={1\over 2}$, the extraction of precise magnetic hyperfine structure constants from the measurements is not hampered by the presence of higher order nuclear moments. The relative size of $\epsilon_{P}$ grows with nuclear charge number $Z$ , and is about 1/3 of $\epsilon_{S}$ in Fr~\cite{stroke61}. 

\begin{figure}
\leavevmode \centering
  \includegraphics[width=2.8in]{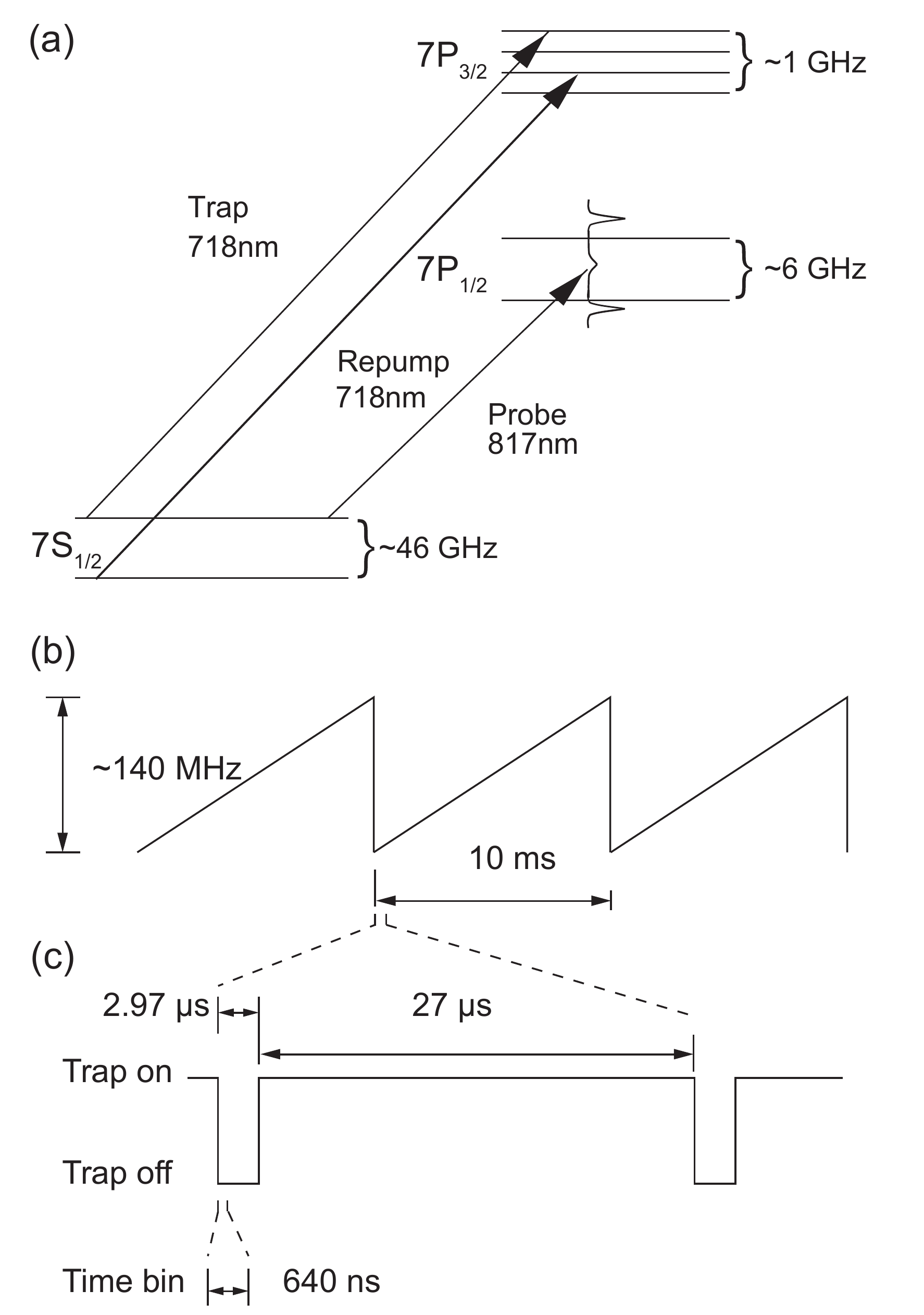}
  \caption{Measurement scheme. (a) Atomic energy levels relevant for trapping and measuring. (b) Sideband frequency scan around 2.9 GHz (isotope-dependent). (c) Time sequence for trapping and data collection. }
  \label{scheme}
\end{figure}

We measure the  $7P_{1/2}$ hyperfine splitting of Fr at the 100 ppm level in a number of isotopes.  We use these measurements in combination with the $7S_{1/2}$ hyperfine splittings \cite{coc85, voss13, voss15, duong87} to determine $R_{\rm HFS}(A)$ to study changes in the hyperfine anomaly. These measurements are carried out at the Francium Trapping Facility (FTF) at TRIUMF \cite{tandecki13a}. We briefly summarize the operation of the FTF: A 500 MeV proton beam irradiates a target that consists of uranium carbide foils to produce between $10^{7}$ to $10^{9}$ Fr$^+$/sec of the selected isotope \cite{bricault14}. We produce an ultracold sample of neutral Fr atoms for Doppler-free spectroscopy by capturing typically a few $10^5$ atoms in a magneto-optical trap (MOT). Two Ti:Sapphire lasers (trap and repumper) form the MOT on the $D_2$ line (718 nm) and leave the $D_1$ line (817 nm) background free for the measurement (Fig. \ref{scheme}(a)). A computer-controlled Fabry-Perot cavity monitors and stabilizes \cite{zhao98} the long-term frequency variation of all of the lasers to better than ${\pm}$5 MHz.

A third Ti:Sapphire laser excites the transition from the upper ground state hyperfine level to each of the two $7P_{1/2}$ hyperfine levels at 817 nm (Fig. \ref{scheme}(a)). We amplitude-modulate the probe laser with a fiber electro-optic modulator (EOM, EOSpace AZ-2K1-10-FPA-FPA-800-UL) that suppresses the carrier and produces sidebands at about half the hyperfine splitting, such that the two sidebands are separated by the size of the splitting  \cite{Note3Carrier}. 
With this method we produce rapid scans of the sidebands \cite{collister14} that minimize many systematic effects compared to scanning the carrier frequency \cite{grossman99}. For instance, we are less sensitive to atom number fluctuations or laser frequency drifts, and the frequency axis of the scan can be precisely characterized since it lies in the microwave regime. We produce the microwaves with phase-locked-loop synthesizers referenced to a Rb clock (SRS FS275). 
The frequency sweep covers 140 MHz in 10 ms (Fig.~\ref{scheme}(b)). The probe beam has a 3 mm diameter with 100 $\mu$W power in each sideband and is retro-reflected to minimize trap displacement from radiation pressure imbalance. 

\begin{figure}
\begin{center}
  \includegraphics[width=0.85\linewidth]{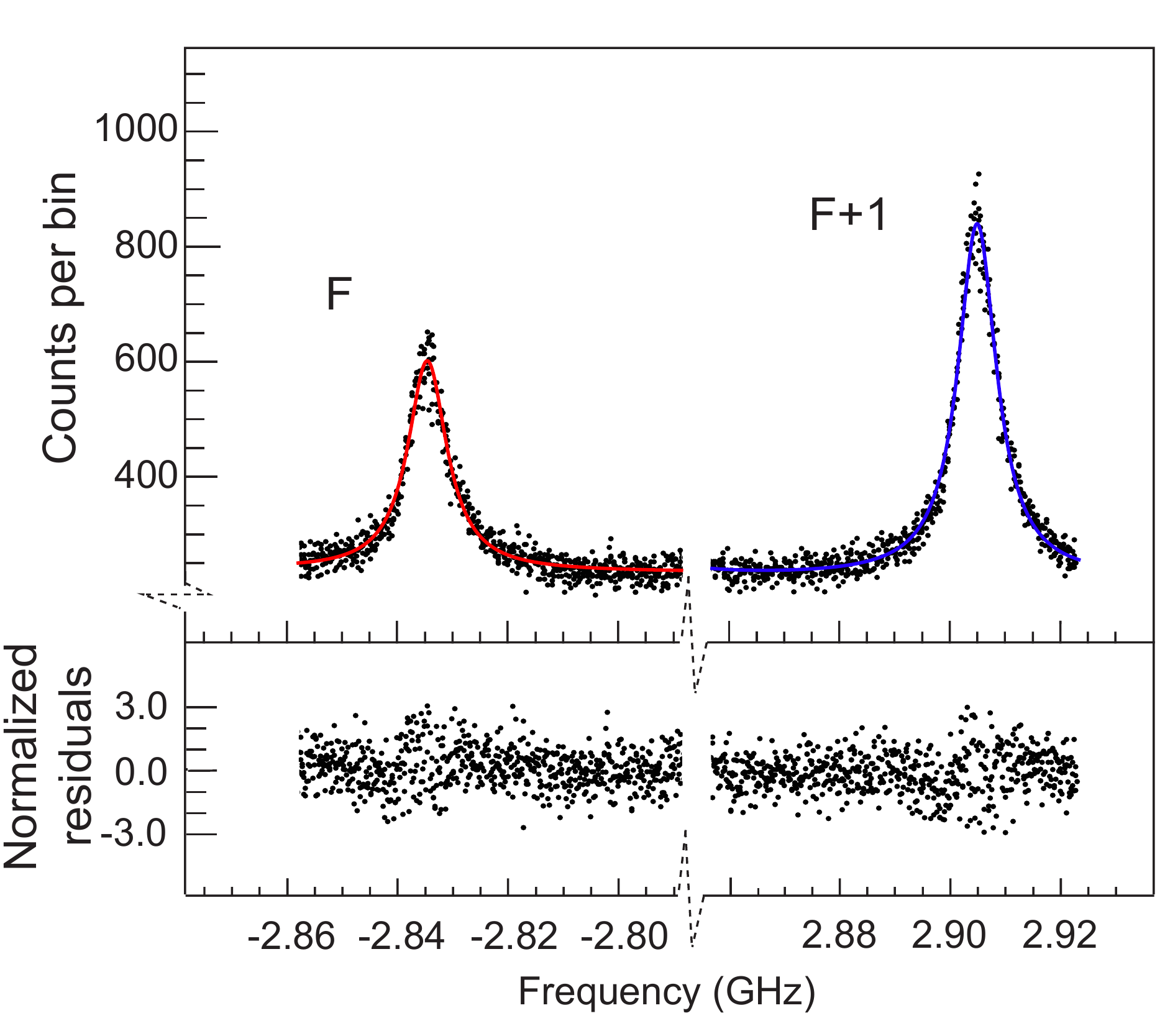}
\caption{Sample of experimental data. Top: Fluorescence counts as a function of sideband frequency for $^{213}$Fr with separate Lorentzian fits to each of the resonances. The hyperfine splitting is the difference of  the frequency position of the two peaks and it gives 5739.60 $\pm$  0.030 MHz. Bottom: Normalized residuals of the fits.}
\label{datafit}
\end{center}
\end{figure}

We collect the fluorescence light with a double relay imaging system (numerical aperture of 0.12) with an interference filter at 817 nm and an edge filter at 795 nm to suppress background light, in particular the trapping laser at 718 nm. A photomultiplier tube (Hamamatsu H7422-50) operating in photon counting mode detects the fluorescence, and we record the photon events as a function of time with a multichannel scaler (SRS SR430). We use a typical bin width of 640 ns with count rates below 250 kilocounts/s. We avoid the ac Stark shift from the MOT trapping light with an experimental cycle of 27 $\mu$s of trapping followed by 2.97 $\mu$s of probing with the trap laser off (Fig.~\ref{scheme}(c)). Data with signal-to-noise ratios of $\gtrsim$ 20 are obtained within a few seconds. Fig.~\ref{datafit} shows a typical spectrum, which yields a HFS splitting with statistical uncertainty at the 30 kHz level. The two peaks indicate the modulation frequency where the +1 (-1) sideband is resonant with the upper (lower) hyperfine peak.

The linear Zeeman effect is the dominant systematic uncertainty, because the quadrupole magnetic field of the MOT stays on during the measurement, and the atoms populate all Zeeman sub-levels.
We characterized this effect by changing the magnetic field gradient (7 to 15 G/cm), probe laser polarization (linear as used in the measurement, to circular), and the position of the atom cloud (1 mm dislocation). We put an upper bound of 540 kHz systematic uncertainty from the Zeeman shift. This agrees well with auxiliary tests in Rb.  
The sideband laser power changes during the scan due to RF power variations, which contributes to a background structure in the data. We take scans without the trapped atoms to record this background. These variations, in addition to line-asymmetries caused by laser drifts during the data accumulation, create small line-shape deviations from a Lorentzian function, as can be seen in the residual structures of the fit in Fig~\ref{datafit}. We evaluate a systematic uncertainty of 100 kHz from the line-shape distortions.
The differential ac Stark shift from the trap laser is mitigated by the fast chopping technique, and the shift from the repump laser is 90 kHz with an uncertainty of 60 kHz \cite{Note2stark}.
Other potential systematic effects include Doppler shifts, probe laser power, frequency calibration and linearity of the scan, which we all evaluated to be at a negligible level. The total systematical uncertainty is 552 kHz. 

\begin{table}
\caption{Isotope, spin ($I$), $7P_{1/2}$ hyperfine splittings (HFS) and ratio $R_{\rm HFS}(A)$ of $7S_{1/2}$ and $7P_{1/2}$ splittings for $\rm ^{206-213, 221}Fr$. We illustrate the neutron orbital configuration ($\nu$ orbital) and its spin alignment with respect to the total nuclear spin ($\nu$ spin). } 
\begin{ruledtabular}
\begin{tabular}{lccccc}
Isotope  & $I$ &  HFS$(7P_{1/2})$ & $R_{\rm HFS}(A)$ & $\nu$ orbital & $\nu$ spin \\ \hline
206$^g$ & 3 &  6009.14(55) & 7.6022(14)$^{b}$ & $a p_{3/2} + b f_{5/2} $ & $\downarrow_{\nu} \uparrow_{I}$ \\
206$^m$ & 7 & 6521.56(57) & 7.6086(10)$^{c}$ & $f_{5/2}$ & $\uparrow_{\nu} \uparrow_{I}$\\
207 & 4.5 &  5559.04(55) & 7.6308(12)$^{d}$ & &\\
208 & 7  & 6561.0(2.3)$^{a}$ & 7.6053(30)$^{c, d}$ & $f_{5/2}$ & $\uparrow_{\nu} \uparrow_{I}$\\
209 & 4.5 & 5639.5(1.0)$^{a}$ & 7.6307(16)$^{d}$ & &\\
209 & 4.5 & 5638.36(56) & 7.6323(13)$^{d}$ & &\\
210 & 6 &  6150.9(1.3)$^{a}$ & 7.6035(17)$^{d}$ & $f_{5/2}$ & $\uparrow_{\nu} \uparrow_{I}$\\
211 & 4.5 & 5710.5(1.0)$^{a}$ & 7.6297(15)$^{d}$ & &\\
212 &  5 & 6556.0(1.0)$^{a}$ & 7.6042(17)$^{d}$ & $p_{1/2}$ & $\uparrow_{\nu} \uparrow_{I}$\\
213 & 4.5 & 5739.43(55) & 7.6292(18)$^{e}$ & &\\
221 & 2.5 & 2431.0(55)       & 7.6581(26)$^{d}$ & &\\
221 & 2.5 & 2433.0(3.9)$^{f}$ & 7.652(12) $^{d}$& & \\
\end{tabular}
\end{ruledtabular}
\label{splittings}
$^{a}$ Ref. \cite{grossman99}. The $7S_{1/2}$ values come from  
$^{b}$ Ref. \cite{voss13},
$^{c}$ Ref. \cite{voss15},
$^{d}$ Ref. \cite{coc85},
$^{e}$ Ref. \cite{duong87}. For $^{208}$Fr,  the HFS$(7S_{1/2})$ is the weighted average and scaled uncertainty of the two measurements following Ref. \cite{pdg}. 
$^{f}$ Ref. \cite{lu97}.
\end{table}

\begin{figure}
\begin{center}
  \includegraphics[width=0.95\linewidth]{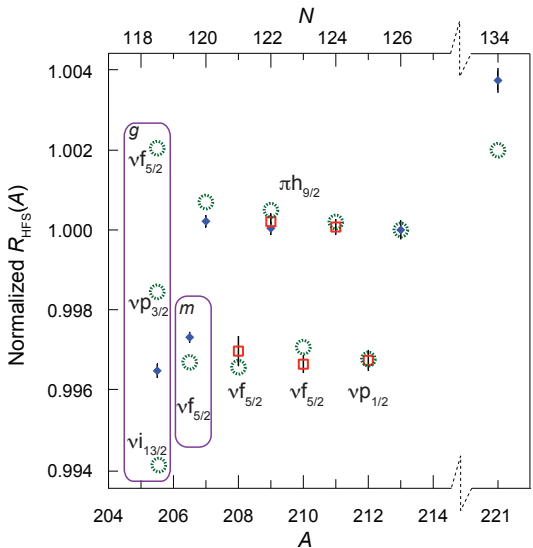}
\caption{ (color online) $R_{\rm HFS}(A)$ as defined in Eq.~2, for several Fr isotopes normalized to the value for $^{213}$Fr (closed neutron shell N=126). Blue diamonds are current measurements. Red squares are from Ref.~\cite{grossman99}. The green circles are predictions based on Ref.~\cite{stroke61}. The ground state $^{206g}$Fr  is shown to the left of the isomer $^{206m}$Fr. Both are enclosed in rectangles with the respective calculations. The proton resides on the indicated $\pi h_{9/2}$ orbital. }
\label{ratios}
\end{center}
\end{figure}

Table~\ref{splittings} shows the hyperfine splittings of the $7P_{1/2}$ state for the five isotopes from this work, together with those reported in Ref.~\cite{grossman99}. The uncertainty includes both the statistical and systematic error contributions stated above. For $^{209}$Fr and $^{221}$Fr we find good agreement with Ref. \cite{grossman99} and Ref. \cite{lu97}, respectively, with smaller error bars.
 The table also lists the ratio $R_{\rm HFS}(A)$ introduced in Eq. \ref{rho}, based on the literature values for the $7S_{1/2}$ splittings \cite{duong87, coc85, voss13, voss15}.
The normalized ratio of the hyperfine anomalies $R_{\rm HFS}(A)/R_{\rm HFS}(213)$ (with $^{213}$Fr taken as the reference isotope for convenience), in a chain of isotopes $A$ = 206-213 and 221 is shown in Fig. \ref{ratios}.
The isotopes span neutron numbers between  $N$ = 119 to 134. The red squares correspond to measurements from Ref.~\cite{grossman99}, and the blue diamonds are the new results. For $A$ = 206 we measured both the low spin ($I$ = 3) nuclear ground state $^{206g}$Fr, and the first high-spin ($I$ = 7), long-lived isomeric state $^{206m}$Fr (lifetime$>$10 s, deduced from MOT lifetime). We clearly distinguish  $^{206m}$Fr from  $^{206g}$Fr in the MOT due to their different trapping and repumping laser frequencies.

The shell model explains reasonably well the magnetic moments of the light Fr isotopes near $N$ = 126~\cite{ekstrom98}. We consider only the dominant orbitals of single nucleons in the shell model to calculate hyperfine anomaly differences for the $7S_{1/2}$ and $7P_{1/2}$ electronic states~\cite{buttgenbach84,stroke61}. The total anomaly combining the proton and the neutron is given by \cite{buttgenbach84}
\begin{equation}
\epsilon_{S} -\epsilon_{P} = \epsilon_{\pi} \beta_{\pi} + \epsilon_{\nu} \beta_{\nu}, \label{anomalysum}
\end{equation}
where $\beta_{\pi,\nu}$ are the fractional contributions to the magnetic moment from the proton and the neutron, respectively. The calculated anomalies (in \%) for a valence proton ($\epsilon_{\pi}$) or neutron ($\epsilon_{\nu}$) for $\pi h_{9/2}$, $\nu p_{1/2}$, $\nu f_{5/2}$, and $\nu p_{3/2}$ orbitals are $-$0.57, $-$3.13,  $-$2.75, and $-$1.75 respectively.  
The total anomaly is represented by the green circles in Fig.~\ref{ratios}. Even though the single particle neutron anomalies are a few times larger than that of the proton, the neutron fractional contribution $\beta_{\nu}$ is typically only 15\%, depending on the orbitals, yielding a contribution from the valence neutron of about the same size as the calculated proton anomaly. 

Our data shows that the even-odd staggering trend is preserved from $^{213}$Fr with a closed neutron shell, down to $^{206}$Fr. The neutron-deficient odd-even isotopes $^{207, 209, 211, 213}$Fr all have near-constant anomalies, differing by a small amount consistent with variations due to the changes in the charge distribution (BR effect) \cite{rosenthal32, crawford49, rosenberg72, martensson95, Note1BR}. This confirms the previous assumption that the anomaly in these isotopes is due primarily to the single valence proton in the $\pi h_{9/2}$ orbital of the spherical shell \cite{grossman99}. $^{221}$Fr is also odd-even, with the valence proton in the same $ \pi h_{9/2}$ orbital. However the 8 neutrons above closed shell create deformation, leading to a rather different anomaly, which we explain in more detail later. For odd-odd isotopes ($^{208,210,212}$Fr), the calculations for the respective neutron orbitals are in good agreement with the experimental data. 

For $^{206g}$Fr, there are some differences compared to a pure $ \nu p_{3/2}$ orbital, which also holds for the magnetic moment~\cite{lynch14,voss13, voss15}. However other nearby orbitals ($\nu f_{5/2}$ or $\nu   i_{13/2}$) give very different values, or even opposite signs with respect to the normalization. The calculations using different orbitals for $^{206g}$Fr are shown in Fig.~\ref{ratios}.  
We note that the odd-odd isotopes have the same sign and roughly the same value of the neutron anomaly contribution. This is a coincidence resulting from the angular momentum coupling, even though their neutron orbitals and nuclear spins are different. This is illustrated by the fact the calculation for lower nuclear spin ($I=3$) $\nu f_{5/2}$ ($^{206g}$Fr) produces an opposite sign compared to the higher spin isotopes ($^{206m, 208, 210}$Fr), as $I=3$ demands the neutron spin to be anti-aligned with the total nuclear spin (see the last column of Table \ref{splittings}).

Deviating from the spherical shell model, the Nilsson picture~\cite{mottelson55} considers the nuclear energy level changes due to a deformation. Calculations~\cite{davidson68} have shown that the $f_{5/2}$ and  $p_{3/2}$ neutron orbitals have a level-crossing of the [521 1/2] and [503 5/2] sub-states at a very small (negative) deformation parameter $|\epsilon_2| \leq 0.05$, a range consistent with the measured quadrupole moment~\cite{voss15}. The valence neutron of $^{206}$Fr can have a mixture of these two orbitals, which shows effects in the magnetic moment~\cite{voss15}. The anomaly is more sensitive to this orbital mixture, and could have a value outside of the range delimited by the two pure orbital predictions. This is because operator evaluations such as $\langle f_{5/2} |r^2 | p_{3/2} \rangle$ break the orthogonality of the two eigenfunctions, resulting a linear dependence of the anomaly to such mixtures, compared to the quadratic dependence of the magnetic moment.

The neutron-rich odd-even isotope $^{221}$Fr has only the proton anomaly contribution. Its Nilsson deformation parameter lies in a positive small to intermediate range,  $0.1 \leq \epsilon_2 \leq 0.2$~\cite{ekstrom86}, and the valence proton still occupies the $h_{9/2}$ orbital. However, its angular momentum projects on to the nuclear symmetry axis, such that the nuclear spin becomes $I=5/2$ \cite{davidson68}. Calculations in \cite{davidson68} yield more than 95\% of the wavefunction in the $\pi h_{9/2}$[523 5/2] state. 
Treating $^{221}$Fr  with this predominant contribution we obtain a correct sign of the anomaly with respect to $^{213}$Fr, as shown in Fig.~\ref{ratios}, and a magnetic moment of  $+1.85~ \mu_{N}$, in reasonable agreement with the value  $+1.58(3)~ \mu_{N}$ empirically scaled from the hyperfine splitting~\cite{ekstrom86}. We note that this is a considerably simpler picture than the nuclear model used in Ref.  \cite{ekstrom86}. The Nilsson calculations are parameter-dependent, and the states with lower angular momentum projections used in Ref. \cite{ekstrom86} would involve more contributions from configuration mixing effects.

In conclusion we present precise measurements of the $7P_{1/2}$ hyperfine splitting in several francium isotopes. The results allow us to study the hyperfine anomaly starting from a closed neutron shell ($^{213}$Fr $N$ = 126) with a simple nuclear distribution, to the boundaries of the single-particle spherical shell model. We demonstrate high-quality spectroscopic measurements, both with ground state nuclei as well as with an isomer. The present results provide the basis for testing the validity and accuracy of future nuclear structure calculations, which will be necessary to extract weak interaction physics from parity non-conservation measurements in francium.

We thank  A. Poves for helpful discussions. We acknowledge support from DOE and NSF from the USA, NRC through TRIUMF and NSERC from Canada, and CONACYT from Mexico. R.C. acknowledges financial support from a University of Manitoba Graduate Fellowship.


\begin{thebibliography}{35}
\bibitem [{Note1()}]{Note1BR}%
  \BibitemOpen
  \bibinfo {note} {The BR effect in the Fr region is at the 10\% level of the
  BW effect, and is model dependent. Our inclusion of this effect is based on the calculation in Ref. \cite{martensson95}.}\BibitemShut {Stop}%
\bibitem [{Note3()}]{Note3Carrier}%
  \BibitemOpen
  \bibinfo {note} {The fiber EOM is a Mach-Zehnder interferometer with the offset point locked by a bias controller (YYlabs Micro-MBC-1) to a null point for suppressing the carrier, which contributes to the background signal. We keep the laser locked near the midpoint of the two hyperfine transitions and scan the modulation frequency. }\BibitemShut {Stop}%
\bibitem [{Note2()}]{Note2stark}%
  \BibitemOpen
  \bibinfo {note} {By comparing data with the trapping beam on and off we
  measure a 5 MHz common shift of the two hyperfine transitions, a 250 $\pm $
  100 kHz differential shift, and a line width increase from 10 to 14
  MHz. We scale this with the repump laser power, taking into account the trap laser detuning,
  and obtain less than 60 kHz differential Stark shift }\BibitemShut {NoStop}%
\end{thebibliography}
\end{document}